\documentclass[aps,pra,twocolumn,showpacs]{revtex4-1}
\usepackage{graphicx,color}
\usepackage{amsmath}
\usepackage{braket}
\usepackage[percent]{overpic}
\usepackage{hyperref}

\newcommand{\bolddell}{{\mbox{\boldmath $d \ell$}}}
\newcommand{\boldr}{{\bf{r}}}
\newcommand{\boldv}{{\bf{v}}}
\newcommand{\ho}{\text{HO}} 
\newcommand{\TF}{\text{TF}} 

\begin{document}

\title[ ]{Vinen turbulence via the decay of multicharged 
vortices in trapped atomic Bose-Einstein condensates}

\author{A.~Cidrim$^1$}
\author{A.~J.~Allen$^2$} 
\author{A.~C.~White$^3$}
\author{V.~S.~Bagnato$^1$}
\author{C.~F.~Barenghi$^2$}

\affiliation{$^1$ Instituto de F\'{\i}sica de S\~{a}o Carlos,
Universidade de S\~{a}o Paulo, C.P. 369,
13560-970 S\~{a}o Carlos, SP, Brazil}

\affiliation{$^2$ Joint Quantum Centre (JQC) Durham-Newcastle,
School of Mathematics and Statistics,
Newcastle University, Newcastle upon Tyne NE1 7RU, UK}

\affiliation{$^3$ Quantum Systems Unit, Okinawa Institute of Science and
	Technology, Okinawa 904-0495, Japan}

\date{\today}

\begin{abstract}
We investigate a procedure to generate turbulence in a trapped
Bose-Einstein condensate which takes advantage of the decay 
of multicharged vortices. We show that the resulting singly-charged vortices 
twist around each other, intertwined in the shape of
helical Kelvin waves, which collide and undergo vortex reconnections, 
creating a disordered vortex state. 
By examining the velocity statistics,
the energy spectrum, the correlation functions and the temporal decay,
and comparing these properties with the properties of ordinary turbulence
and observations in superfluid helium, we conclude that this disordered
vortex state can be identified with the `Vinen' regime of
turbulence which has been discovered in the context of
superfluid helium.
\end{abstract}


\maketitle

\section{Motivation}

The singular nature of quantized vorticity (concentrated along vortex
lines) and the absence of viscosity make superfluids remarkably
different from ordinary fluids. Nevertheless, recent 
studies~\cite{Barenghi2014b} have revealed that 
superfluid helium, when suitably stirred, shares an 
important property with ordinary turbulence:
the same Kolmogorov energy spectrum~\cite{Frisch}, describing a 
distribution of kinetic energy over the length scales which signifies 
an energy cascade from large 
length scales to small length scales. This finding suggests that turbulence of quantized vortices (quantum turbulence) 
may represent the `skeleton' 
of ordinary (classical) turbulence~\cite{Hanninen2014}.  
A puzzle arises however: experiments \cite{Walmsley2008a,Barenghi2016} 
show that there are other regimes in which turbulent superfluid
helium lacks the Kolmogorov spectrum.

Trapped atomic Bose-Einstein condensates (BECs) are ideal systems 
to tackle this puzzle.
This is because, unlike
helium, the physical properties of BECs (e.g. the strength
of atomic interactions, the density, the vortex core radius) can be
controlled.  Moreover, individual quantized vortices 
are more easily nucleated, manipulated~\cite{Aioi2011,Davis2009}
and observed~\cite{Madison2000,Raman2001,Freilich2010} 
in BECs than in helium. 

A second motivation behind our work is that
the study of three-dimensional turbulence in BECs
\cite{Tsatsos2016} is held back by the lack of 
a standard method to excite turbulence 
in a reproducible way, so that experiments 
and numerical simulations can be compared with each other and 
any generality of results can be more easily recognized.
In classical turbulence, standard benchmarks are  
flows driven along channels or stirred by propellers,
flows around well-defined obstacles
(e.g. cylinders, spheres, steps) and wind tunnel flows past grids. 
Similar techniques are used for superfluid helium 
\cite{Barenghi2014a,Skrbek-Sreeni2012}, 
which is mechanically
or thermally driven along channels, or stirred  
by oscillating wires, grids, forks, propellers and spheres.
Vortices and turbulence in BECs have been generated by 
moving a laser beam across the  
BEC~\cite{Raman2001,Neely2010,White2012,White2014,
Kwon2014,Stagg2015,Allen2014b,Cidrim2016}, 
by shaking~\cite{Henn2009} or stirring the trap 
rotating it around two perpendicular axes~\cite{Kobayashi2007}, by phase 
imprinting staggered vortices~\cite{White2010}, or by thermally 
quenching the system (Kibble-Zurek mechanism)~\cite{Weiler2008,Chomaz2015,
Lamporesi2014,Navon2015}. This variety of techniques
and the arbitrarily chosen values
of physical parameters means that comparisons are difficult. 
Moreover, the disadvantage of some of these techniques is that they 
tend to induce 
large surface oscillations or even fragmentation~\cite{Parker2005}
of the condensate which complicate the interpretation of results 
and the comparison with classical turbulence.

The aim of this report is to propose  a technique to induce 
turbulence in BECs based on the decay of multicharged 
vortices (a more controlled and less forceful technique than the above 
mentioned methods), and to characterize the turbulence which is produced.
For this purpose, we shall compare two disordered
vortex states, which we shall call `anisotropic' and `quasi-isotropic' for simplicity,
resulting respectively from the decay of a single $j=4$ charged vortex
and from the decay of two antiparallel $j=2$ vortices.
We shall reveal a new connection between BEC turbulence and the so-called `Vinen' turbulent regime discovered in superfluid helium.

\section{Multicharged vortices}

In a superfluid, the circulation around a vortex
line is an integer multiple ($j=1,2,\cdots$) of the quantum of circulation
\cite{Primer} $\kappa=h/m$, that is
$\oint_C \boldv \cdot \bolddell=j \kappa$
where $C$ is a closed path around the vortex line,
$h$ is Planck's constant and $m$ the atomic mass.
The velocity field around an isolated vortex line is therefore constrained
to the form $v= j \kappa/(2 \pi r)$ where $r$ is the distance to the vortex
axis. This property is in marked contrast to ordinary fluids where the velocity
of rotation about an axis (e.g. swirls, tornadoes, galaxies)
has arbitrary strength and radial dependence.

The angular momentum and the energy of an isolated vortex in a homogeneous
superfluid grow respectively with $j$ and $j^2$~\cite{Primer}. 
Therefore, for the same angular momentum, multicharged ($j>1$) 
vortices carry more energy, and, in the presence of thermal dissipative 
mechanisms, tend to decay into singly-charged 
vortices~\cite{Shin2004,Kumakura2006,Okano2006,Isoshima2007}. 
Besides the energy instability, there is also a dynamical 
instability~\cite{Pu1999,Mottonen2003,Huhtamaki2006a} 
which would destabilize a multicharged vortex. The time-scale of
these effects has been investigated 
\cite{Mottonen2003,Shin2004,Huhtamaki2006b,Mateo2006,Okano2006,Isoshima2007,Kuwamoto2010,Kuopanportti2010}.
The technique of topological phase imprinting \cite{Nakahara2000}
has allowed the controlled generation of multicharged vortices
\cite{Leanhardt2002,Shin2004,Okano2006,Kuwamoto2010} in atomic condensates. 
The decay of a doubly-quantized vortex into two singly-quantized 
vortices has been studied~\cite{Shin2004,Huhtamaki2006b,Mateo2006} in a
Na BEC.  Quadruply-charged quantized vortices are also theoretically 
predicted~\cite{Kawaguchi2004} to decay. Recent work has determined that the
stability of such vortices is affected by the condensate's density \cite{Shin2004} and size \cite{Mateo2006}, and by the nature of the perturbations \cite{Kawaguchi2004}.

\section{Model}

We model the condensate's dynamics using the 3D Gross-Pitaevskii 
equation (GPE) \cite{Primer} for a zero-temperature condensate
\begin{equation}
       i \hbar \frac {\partial \psi}{\partial t} = 
       \left( 
       -\frac{\hbar ^2}{2m}\nabla^2 + U(\mathbf{r}) + g\vert \psi \vert^2  
       \right) \psi,
\label{eq:gpe}
\end{equation}
\noindent
where $\psi(\boldr,t)$ is the condensate's wavefunction, 
$\boldr$ the position, $t$ the time, and 
\begin{equation}
U(\boldr)=\frac{m}{2}(\omega_x^2 x^2 + \omega_y^2 y^2 + \omega_z^2 z^2)=
\frac{m}{2}(\omega_r r^2+\omega_z z^2),
\end{equation}
\noindent
the harmonic trapping potential. The parameter
$g=4\pi \hbar^2 a_s/m$ characterizes
the strength of the inter-atomic interactions, 
$a_s$ is the s-wave scattering length. The normalization is $\int_V \vert \psi \vert^2 dV=N$ where
$V$ is the BEC's volume and $N$ the number of atoms. 
We cast the GPE in dimensionless form using 
$\tau_{\ho}=\omega_r^{-1}$, $\ell_{\ho}=\sqrt{\hbar/m\omega_r}$
and $\hbar\omega_r$ as units of time, distance and energy respectively.
The inter-atomic interaction parameter $g$ is
chosen to describe a typical BEC with $N\approx 1\times 10^{5}$ atoms 
of $^{87}$Rb trapped harmonically in a cigar-shaped BEC with
radial and axial frequencies such that
$\omega_z/\omega_r=\lambda=0.129$. It is of our particular interest to study properties of the condensate's velocity field components, which are computed from the definition $\mathbf{v}(\mathbf{r})=\left(\psi\nabla\psi^{*}-\psi^{*}\nabla\psi\right)/2i\left|\psi\right|^2$. The dimensionless GPE is solved numerically in the 3D domain 
$-10~\ell_{\ho}\leq x,y \leq 10~\ell_{\ho}$ and 
$-40~\ell_{\ho}\leq z \leq 40~\ell_{\ho}$ 
on a $128\times128\times512$ grid 
(keeping the same spatial discretization in the three directions) 
with time-step $\Delta t = 10^{-3}$ using the 
4th order Runge-Kutta method with XMDS \cite{Dennis2013}. 
We have performed tests with different grid sizes and verified that our 
results are independent of the discretization. 

\begin{figure}[t!]
	\centering{\includegraphics[scale = 0.12]{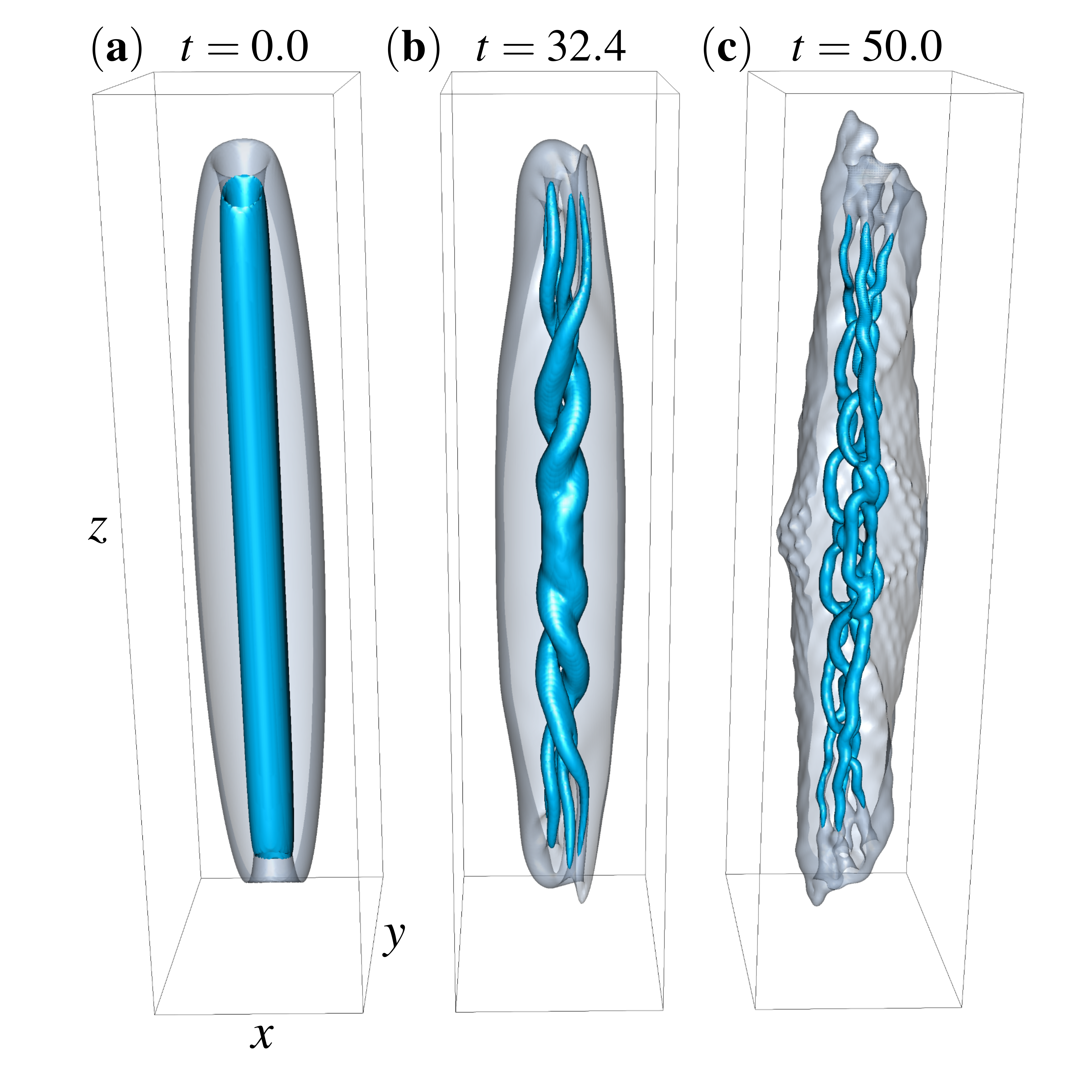}}
	\caption{
		(Color online).
		The 3D isodensity plots show the time (in units of $\tau_\mathrm{HO}$) evolution of an initial
		$j=4$ multicharged vortex (a); notice the twisted unwinding of the
		vortex (b), which finally decays into four singly-charged ($j=1$) 
		vortices (c) in the disordered, anisotropic state.
		The density isosurfaces representing the condensate
		and the vortices are respectively grey and blue. The axes directions are labeled as $x$, $y$, and $z$, in units of $\ell_\mathrm{HO}$.
	}
	\label{fig1}
\end{figure}

\section{Decay of single quadruply-charged vortex}

The shape of the singly-charged vortex lines emerging from the decay of a
multicharged vortex depends on where and when the decay starts. 
The singly-charged lines
may be straight or intertwined (as reported here)
depending on the perturbation's symmetry and the local density homogeneity.
If the perturbation is uniform and the density does not
vary much in the $z$-direction, every point on the vortex unwinds
at the same rate, and  straight singly-charged vortex lines 
will emerge.  However, if the density changes significantly 
along $z$, the unwinding takes place at different times at different
positions, inducing intertwining, as discussed in~\cite{Huhtamaki2006b}.

The twisted vortex decay is shown in Fig.~\ref{fig1} and in the first 
movie in the Supplementary Material~\cite{SuppMovie}. The main feature which is visible during the evolution
are Kelvin waves. Kelvin waves consist of
helical displacements of the vortex core axis, and play an important role
in quantum turbulence \cite{Barenghi2014a}; they have been recently
experimentally identified in superfluid helium \cite{Fonda2014} and their
presence has been recognized in atomic condensates \cite{Bretin03,Smith04}.
Considering Fig.~\ref{fig1}, 
it is worth distinguishing the Kelvin waves which, in our case, 
emerge on parallel vortices from the decay of a multicharged 
vortex~\cite{Mottonen2003,Mateo2006} in a confined geometry,
from the Kelvin waves generated by the Crow instability~\cite{Simula2011}
on anti-parallel vortices in a homogeneous condensate.

Our numerical experiments suggest that the decay  
of the multicharged vortex can be sped up. Imposing 
random fluctuations ($\leq 10\%$ of $\vert \psi \vert$) to the 
initial $j=4$ wave function does not significantly change the decay time 
scale, probably because the symmetry of the initial condition is not 
completely broken.  A small displacement of the vortex core axis 
($\approx 0.5~\ell_{\ho}$) is more efficient, triggering the onset of 
the twisted unwinding in about 2.4$~\tau_{\ho}$; 
a larger displacement ($\approx  0.2~\ell_{\ho}$) reduces this time 
to 2.0$~\tau_{\ho}$. Among the other methods which we have
investigated, the most efficient is to gently squeeze the harmonic 
potential in the $xy$ plane by an amount $\omega_x/\omega_y=0.9$ 
when preparing the initial state in imaginary time, then resetting 
$\omega_x/\omega_y=1$ when propagating the GPE in real time; 
the squeeze triggers the onset of decay in only 
1.0$~\tau_{\ho}$. In the experiments, it is usually
difficult to control perturbations well enough to reproducibly 
determine the time scale of decay. 

\begin{figure}[t!]
	\centering{\includegraphics[scale = 0.43]{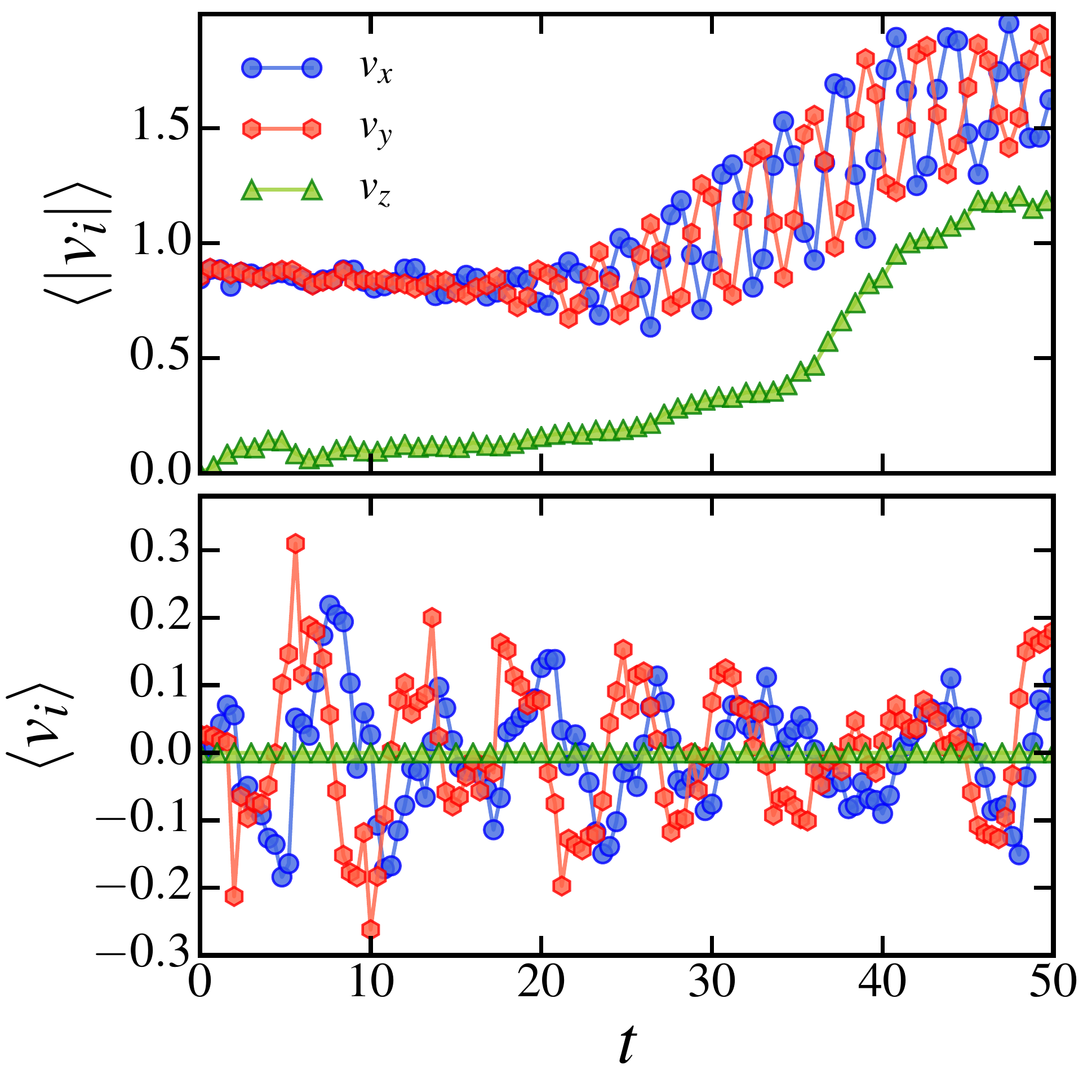}}\\
	\caption{
		(Color online). 
		Decay of quadruply-charged vortex (anisotropic state).
		Time evolution of average velocity components $v_i$ and $\vert v_i \vert$
		(in units of $\ell_{ho}/\tau_{\ho}$, where $i=x,y,z$)
		vs time (in units of $\tau_{\ho}$). The symbol $\langle \cdots \rangle$
		denotes the spatial average over the condensate.}
	\label{fig2}
\end{figure}

Fig.~\ref{fig2} shows the time evolution of the (spatially) averaged velocity 
components and their magnitudes during the decay.  The $x$- and $y$-components
display oscillations which become large for $t>20~\tau_{\ho}$, 
after the initial multicharged vortex has split. 
The $z$-component behaves differently
because all vortices are aligned in the $z$-direction. For simplicity of reference, we call an `anisotropic state' the resulting disordered vortex configuration.

\begin{figure}
	\centering{\includegraphics[scale = 0.58]{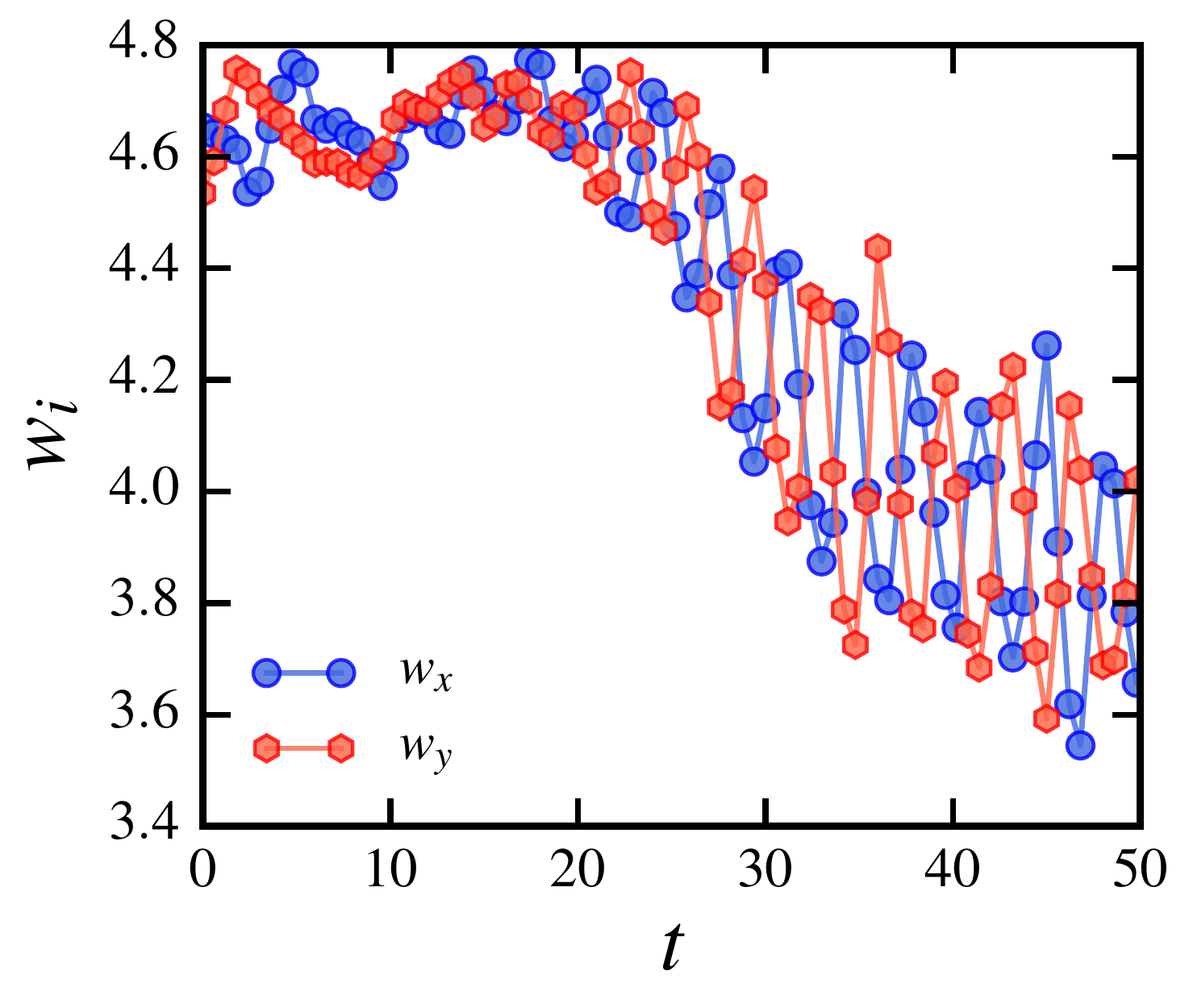}}\\
	\caption{
		(Color online). 
		Decay of quadruply-charged vortex (anisotropic state). Time (in units of $\tau_\ho$) evolution of the condensate's width $w_i$ (in units of $\ell_\ho$).
	}
	\label{fig3}
\end{figure}

\section{Decay of two antiparallel doubly-charged vortices}

\begin{figure}[t!]
	\centering{\includegraphics[scale = 0.12]{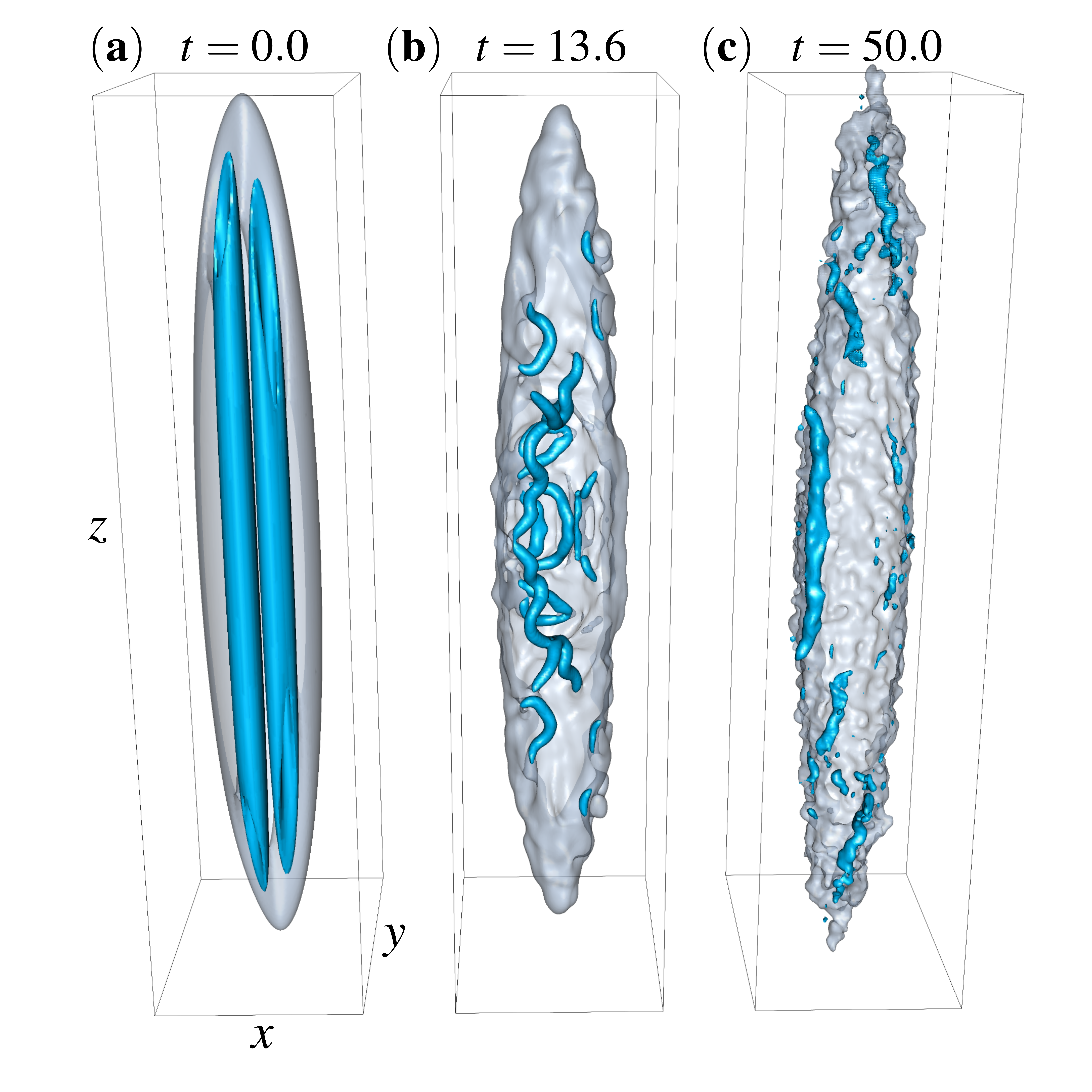}}\\
	\caption{
		(Color online). 
		Isodensity plots showing the evolution (with time $t$ in units of $\tau_{\ho}$)
		of two initial doubly-charged ($j=2$) antiparallel vortices at $t=0~\tau_{\ho}$ (a)
		into the turbulent quasi-isotropic state at
		$t=13.6~\tau_{\ho}$ (b), which has finally decayed at $t=50~\tau_{\ho}$ (c). The axes directions are labeled as $x$, $y$, and $z$, in units of $\ell_\mathrm{HO}$.}
	\label{fig4}
\end{figure}

Now we exploit the twisted unwinding of a multicharged 
vortex as a convenient technique to generate turbulent vortex tangles which are
relatively free of large density perturbations.
We start by numerically imprinting
anti-parallel, doubly-charged vortices 
as the initial state. One vortex is centered at position 
$(x,y)=(1.8~\ell_{\ho}, 1.5~\ell_{\ho})$ and the other 
at $(x,y) = (1.0~\ell_{\ho}, -1.3~\ell_{\ho})$, as shown 
in Fig.~\ref{fig4}(a). During the evolution,
the vortices unwind, twist, move slightly 
forward due to the self-induced velocity field, and then reconnect, 
generating a turbulent state with only moderate density 
oscillations, as shown in Fig.~\ref{fig4}(c).

\begin{figure}
	\centering{\includegraphics[scale = 0.43]{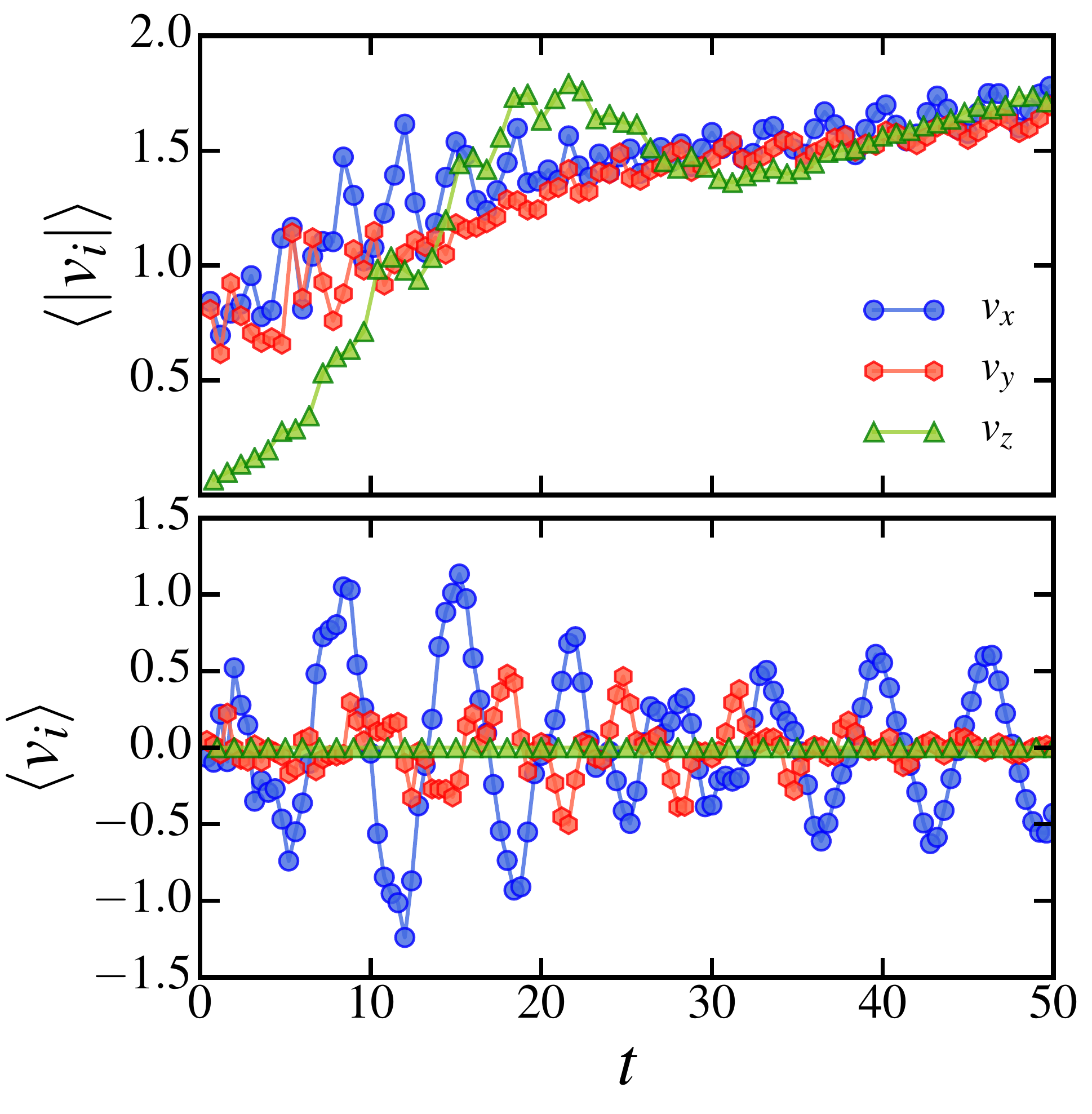}}\\
	\caption{
		(Color online). 
		Decay of antiparallel doubly-charged
		vortices (quasi-isotropic state). 
		Time evolution of velocity components as in Fig.~\ref{fig2}.}
	\label{fig5}
\end{figure}

We analyze the turbulent state in terms of the spatial averages of the 
velocity components. At the beginning of the decay, we find
$\langle \vert v_y \vert \rangle / \langle \vert v_x \vert \rangle \approx 1$
and (as expected, as vortices are initially
aligned in the $z$-direction)
$\langle \vert v_z \vert \rangle/\langle \vert v_x \vert \rangle \approx 0$.
Fig.~\ref{fig5} (top) shows that, as the time proceeds,
the vortex configuration becomes almost isotropic, indeed at $t=10~\tau_{\ho}$
we have
$\langle \vert v_y \vert \rangle/\langle \vert v_x \vert \rangle \approx 1.00$
and 
$\langle \vert v_z \vert \rangle/\langle \vert v_x \vert \rangle \approx 0.77$. For simplicity of
reference, we call a `quasi-isotropic state' the resulting disordered vortex configuration.

\begin{figure}
	\centering{\includegraphics[scale = 0.58]{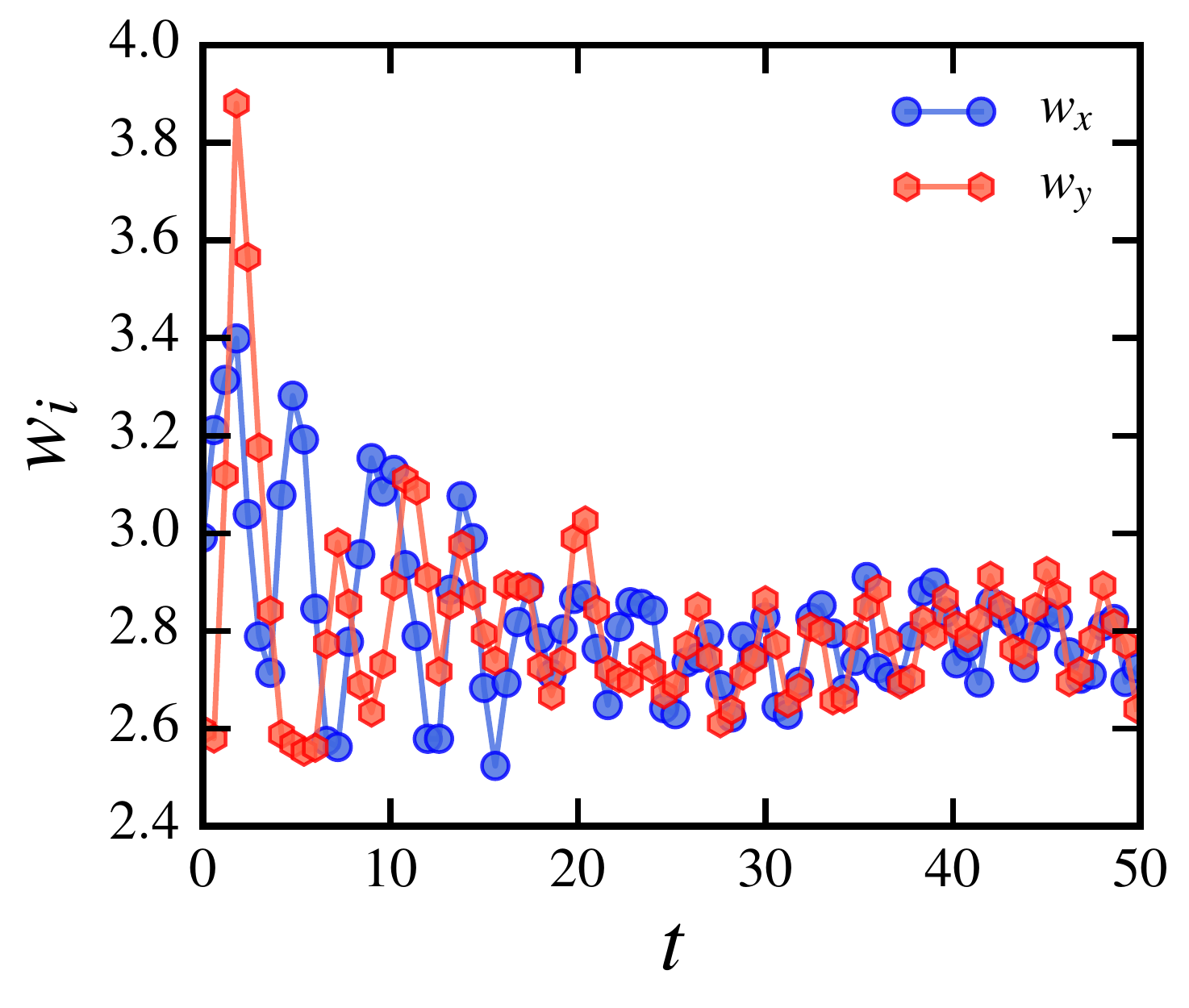}}\\
	\caption{
		(Color online). 
		Decay of antiparallel doubly-charged vortices (quasi-isotropic state). Time evolution of the condensate's width.
	}
	\label{fig6}
\end{figure}

The oscillations of the transverse velocity
components $v_x$ and $v_y$ appear practically in phase for the anisotropic case (see Fig.~\ref{fig2} (bottom))
and almost completely out of phase for the quasi-isotropic case (see Fig.~\ref{fig5} (bottom)), and these suggest the existence of collective modes. In order to properly identify these modes we have evaluated the time evolution of the condensate's transverse widths $w_x$ and $w_y$ (found by adjusting Gaussian fits on $x$- and $y$-directions over the $z$-integrated density) for both cases. As can be seen in Fig.~\ref{fig3}, the anisotropic case exhibits a quadrupolar mode, in which $w_x$ and $w_y$ oscillate out of phase in time. After performing a Fourier analysis of these quadrupolar oscillations we verified the mode's frequency to be $\omega\sim \sqrt{2}\omega_r$, in agreement with theoretical predictions for a vortex-free cloud \cite{Stringari1996}. This mode is reminiscent of the unstable (quadrupolar) Bogoliubov mode that drives the decay of the initial multicharged vortex, as identified for an analogous 2D trapped system in \cite{Kuopanportti2010}. However, in the quasi-isotropic case, Fig.~\ref{fig6} shows that after $t\approx 10~\tau_\ho$ there is a small ($\sim 0.2~\ell_\ho$) in-phase oscillation of the widths (as opposed to the larger oscillations for the anisotropic case in Fig.~\ref{fig3}). The latter can be identified as a breathing mode and (again, through a Fourier analysis) it was found to exhibit a characteristic frequency of $\omega\sim 2\omega_r$, also in agreement with theoretical predictions for a vortex-free cloud \cite{Stringari1996}; this value is expected to hold even for rapidly rotating trapped systems \cite{Watanabe2006}. These results, alongside a visual comparison of Figs.~\ref{fig1} and \ref{fig4}, show that the outer surface of the condensate is actually slightly less disturbed in the quasi-isotropic case. Our turbulent condensate (Fig.~\ref{fig3}(c)) is clearly less `wobbly' than condensates made turbulent via other stirring methods, notably Refs.~\cite{White2010} and \cite{Parker2005}).

\begin{figure}
	\centering{\includegraphics[scale = 0.43]{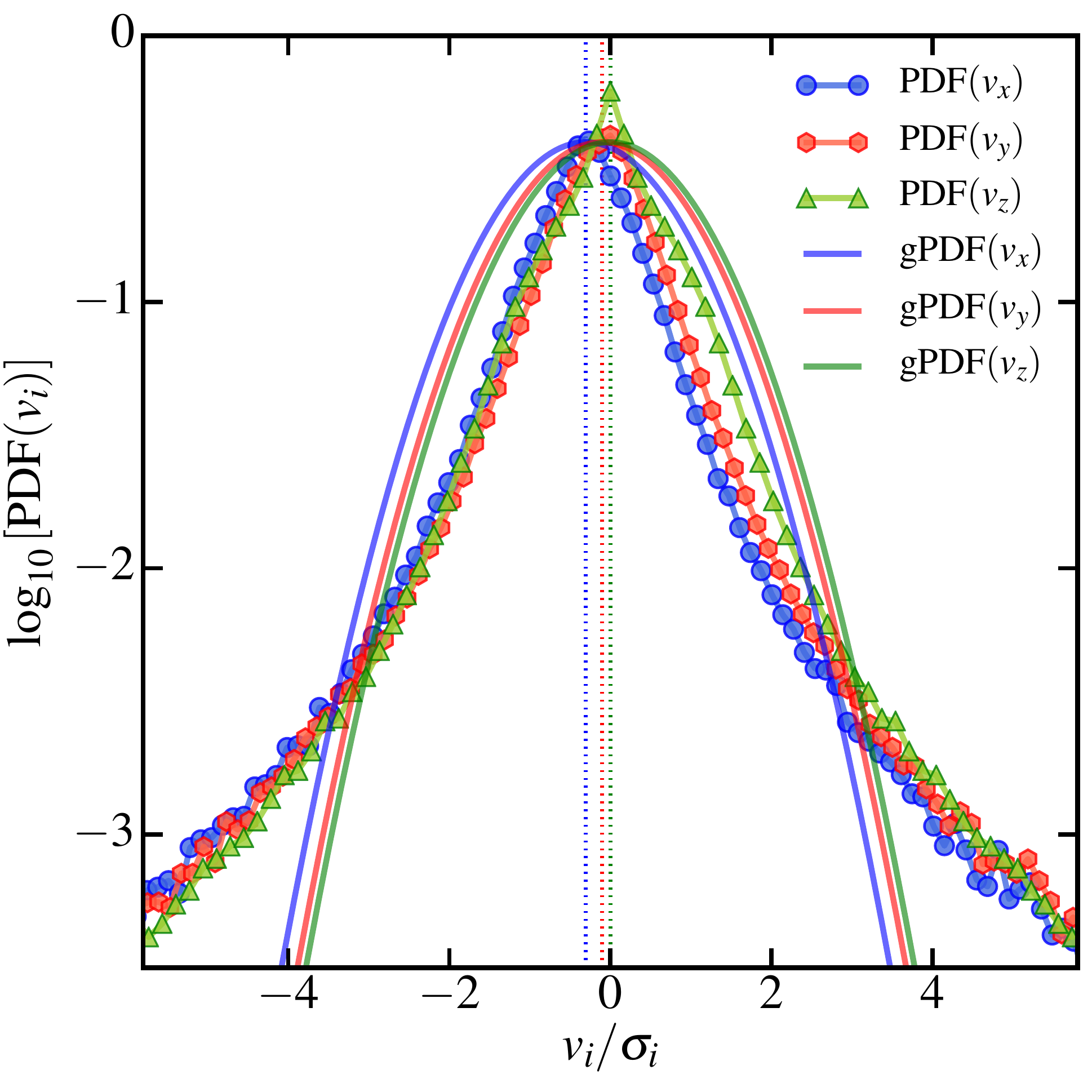}}\\
	\caption{
		(Color online). 
		Decay of antiparallel doubly-charged vortices (quasi-isotropic state).
		PDFs of velocity components $v_i$ ($i=x,y,z$ corresponding to blue, red, 
		and green symbols respectively) plotted vs $v_i/\sigma_i$
		(where $\sigma_i$ are the corresponding standard deviations)
		at $t=12.6~\tau_{\ho}$. 
		For reference, the solid curves are Gaussian fits (gPDFs) with
		standard deviations $\sigma_x=1.8$, $\sigma_y=1.7$ and $\sigma_z=1.4$, and 
		mean values (plotted as vertical lines near the origin)
		$\mu_x=-0.3$, $\mu_y=-0.1$ and $\mu_z=0.0$.
	}
	\label{fig7}
\end{figure}

We proceed and analyze the distribution of values of
the turbulent velocity components (Fig.~\ref{fig7}). We find that the
probability density functions (PDFs, or normalized histograms) 
display the typical power-law scaling
${\rm PDF}(v_i) \sim v_i^{\alpha_i}$
($i=x,y,z$) where $\alpha_x \approx -2.97$, $\alpha_y \approx -2.95$, 
$\alpha_z \approx -3.20$, in agreement with findings in larger
condensates \cite{White2010}. Such power-law scaling,
characteristic of quantum turbulence and 
observed in helium experiments \cite{Paoletti2008}, 
contrasts the Gaussian PDFs which are typical of ordinary turbulence.
The difference between power-law and Gaussian statistics is important
at high velocities (power-law PDFs have larger values in the tails)
and arises from the quantization of vorticity
(which creates very large velocity $v=\kappa/(2 \pi r)$
for $r \to 0$). 
For the sake of comparison, Fig.~\ref{fig7} 
also displays Gaussian fits \cite{gPDF}.

\section{Identification of the turbulence}

The two disordered vortex states, anisotropic and quasi-isotropic, produced 
respectively by the decay of a single quadruply-charged
vortex (Section IV) and by decay of two antiparallel doubly-charged
vortices (Section V) are clearly different. 
In the anisotropic state, all vortex lines are aligned in the same direction, and the net nonzero angular momentum constrains the flow.
In the quasi-isotropic state, the oscillations of the transverse velocity components are three times larger, 
suggesting the presence of high-velocity events (vortex 
reconnections between Kelvin waves growing on opposite-oriented vortices)
which are the hallmarks of turbulence.
Moreover, in the quasi-isotropic state, the zero angular momentum of the initial 
configuration allows a redistribution of the
velocity field, making the amplitudes of the three velocity
components almost equal; indeed, after the initial vortex has split
($t \approx 10~\tau_{\ho}$), the axial velocity 
$\langle \vert v_z \vert \rangle \approx 0.9~\ell_{\ho}/\tau_{\ho}$, 
is not much smaller than the transverse velocity
$\langle \vert v_x \vert \rangle \approx 1.2~\ell_{\ho}/\tau_{\ho}$ 
and $\langle \vert v_y\vert \rangle \approx~1.1~\ell_{\ho}/\tau_{\ho}$.
On the contrary, in the anisotropic state, at the same stage 
($t \approx 10~\tau_{\ho}$), the axial velocity component 
is much smaller than the transverse components.
In other words, the velocity field which results from the decay of the antiparallel doubly-charged vortex state is indeed almost isotropic.

Since there is not yet a precise definition of turbulence, in principle both disordered states investigated here could be considered somewhat `turbulent'. However at this early stage of investigation, we want to make conceptual connections to the simple isotropic cases known in the literature (in particular, the Kolmogorov and Vinen types of turbulence). Therefore hereafter we concentrate on the quasi-isotropic state. 

The next question is whether
the quasi-isotropic state is `turbulent' in the sense of ordinary
turbulence or in comparison with turbulent superfluid helium.
To find the answer, we turn to the workhorse of statistical physics:
the correlation function.

\begin{figure}[t!]
	\centering{\includegraphics[scale = 0.5]{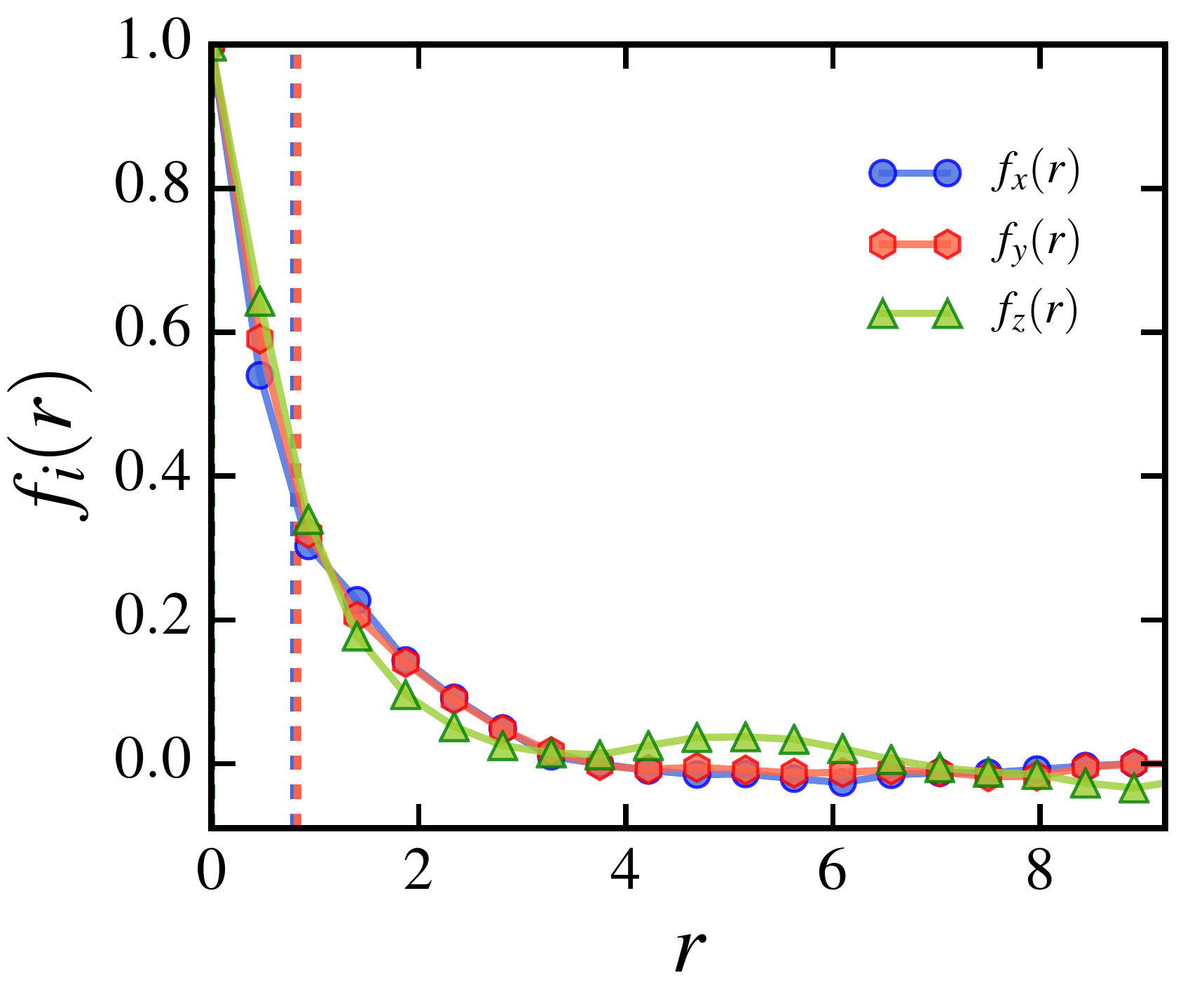}}\\
	\caption{
		(Color online). 
		Decay of antiparallel doubly-charged vortices
		(quasi-isotropic state). Longitudinal correlation functions $f_i(r)$ 
		($i=x,y,z$) vs $r$ (in units of $\ell_{\ho}$) at $t=13~ \tau_{\ho}$.
	    The vertical dashed lines indicate the length $\ell_c$ over which the velocity field is highly correlated, as defined in Eq.(\ref{ellc}).
	}
	\label{fig8}
\end{figure}

The quantity which measures the degree of randomness of a turbulent flow 
is the (normalized) longitudinal correlation 
functions \cite{Stagg2016,Davidson2004}, defined by
\begin{equation}
\label{correlation}
f_i(r)=\frac{\left<v_i(\mathbf{x}+r{{\hat{\bf e}}_i})v_i(\mathbf{x})\right>}{\left<v_i(\mathbf{x})^2\right>},
\end{equation}
\noindent
where the symbol $\langle \cdots \rangle$ denotes average
over $\mathbf{x}$, and ${\hat{\bf e}}_i$ is the unit vector 
in the direction $i=x,y,z$.  
The distance $r$ is limited by the size of the BEC, approximately
the transverse and axial Thomas-Fermi radii $d_{\TF}=4.2~\ell_{\ho}$ 
and $D_{\TF}= 32.6~\ell_{\ho}$.
From the longitudinal correlation function one obtains the integral 
length scale 
\begin{equation}\label{ellc}
\ell_c = \int_{0}^{\infty} f_i(r)dr.
\end{equation}
\noindent
In fluid dynamics, $\ell_c$ represents
the size of the large eddies. In our case, $\ell_c$ is 
the length scale over which the velocity field is highly correlated. 
Fig.~\ref{fig8} shows that the correlation functions
drop to only $\approx 10 \%$ at distances of the order of the
average separation between the vortex lines,
$\ell \approx 6.1~\ell_{\ho}$; the last quantity is
estimated as $\ell \approx L^{-1/2}$ from the measurement of the
vortex line density $L$ (the vortex length per unit volume).
Physically, this lack of correlation means that
the vortex lines are randomly oriented with respect to each other.
The analysis of the correlation function therefore suggests
that the turbulent velocity field arising from the 
decay of antiparallel doubly-charged vortices is essentially a random flow.

\begin{figure}[t!]
	\centering{\includegraphics[scale = 0.38]{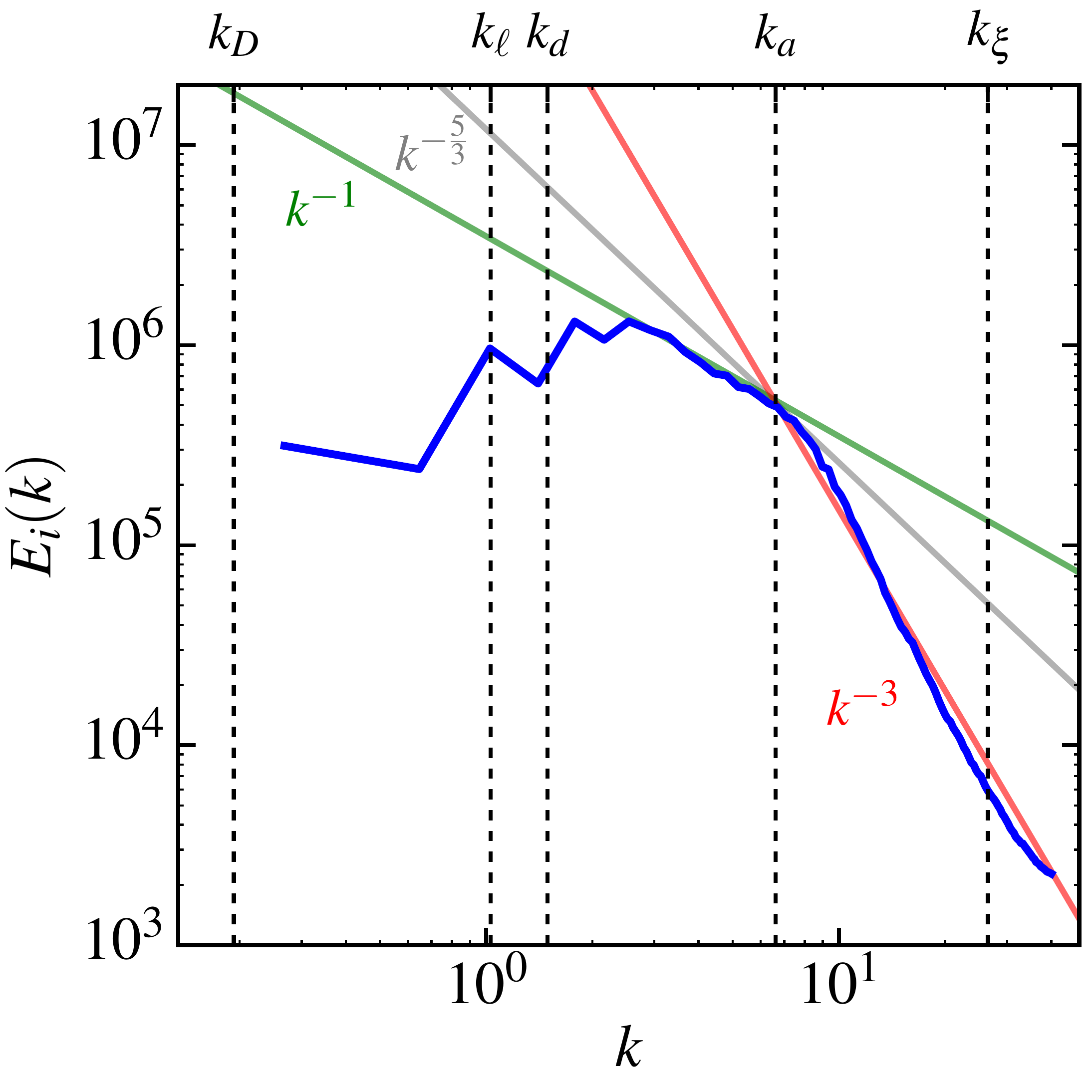}}\\
	\caption{
		(Color online). 
		Decay of two antiparallel doubly-charged vortices (quasi-isotropic state).
		Incompressible kinetic energy spectrum $E_i(k)$ (arbitrary units)
		vs wavenumber $k$ (in units of $\ell_{\ho}^{-1}$)	at $t=12.8~\tau_{\ho}$. The vertical lines mark the wavenumbers corresponding to
		the healing length $\xi=0.24~\ell_{\ho}$, 
		the vortex core size $a=0.96~\ell_{\ho}$,
		the average distance between vortex lines $\ell_v=6.10~ \ell_{\ho}$,
		the radial and axial Thomas-Fermi radii $d_{\TF}=4.21~\ell_{\ho}$
		and $D_{\TF}=32.64~\ell_{\ho}$. Red, grey and green lines mark the 
		power laws $k^{-3}$, $k^{-5/3}$ and $k^{-1}$ respectively.}
	\label{fig9}
\end{figure}

This result implies that the distribution of the kinetic energy over the
length scales, or energy spectrum $E(k)$ (where the wavenumber $k$ 
represents the inverse length scale), should be very different from the
celebrated Kolmogorov scaling, $E(k) \sim k^{-5/3}$, which is observed
in classical turbulence and implies a particular structure of the flow. The importance of the Kolmogorov scaling is 
that it is the signature of a nonlinear cascade mechanism which transfers energy from large length scales to small length scales.  Besides ordinary turbulence, the classical Kolmogorov scaling has been observed in helium experiments \cite{Tabeling1998,Salort2010} when
the turbulence is generated by grids in wind tunnels or 
by counter-rotating propellers. Numerical simulations
\cite{Laurie2012} revealed that the
Kolmogorov scaling is associated with the presence of large-scale
polarization (bundles) of
vortex lines which become locally parallel to each other; the bundles 
locally create a net average rotation over length scales larger 
than $\ell$, thus building up energy at wavenumbers 
$k < k_{\ell}=2 \pi/\ell$ (the hydrodynamical range).


To compute the energy spectrum $E(k)$ of our turbulent condensate, we use 
a standard procedure \cite{Nore1997} to extract the incompressible kinetic
energy from the total energy,  obtaining Fig.~\ref{fig9}.
To interpret the figure, we mark with vertical lines the wavenumbers 
$k_{\xi}=2 \pi /\xi$, $k_a=2 \pi/a$, $k_{\ell}=2 \pi/\ell$,
$k_d=2 \pi/d_{\TF}$ and $k_D=2 \pi/D_{\TF}$
corresponding to the healing length $\xi$, the vortex
core radius $a$,  the average vortex separation $\ell$, and the 
radial and axial Thomas-Fermi radii, $d_{\TF}$ and
$D_{\TF}$. Fig.~\ref{fig9} clearly shows that, at the hydrodynamical
length scales $k<k_{\ell}$ the energy spectrum is not of Kolmogorov type:
under the classical $k^{-5/3}$ scaling, substantially more energy would be 
contained in the small $k$ region. Instead, over a wide range of wavenumbers
up to $k_a$, we observe the $E(k) \sim k^{-1}$
spectrum which is characteristic of an isolated straight vortex line.
This means that, at distances less than $\ell$, 
the velocity field is dominated
by the nearest vortex in the vicinity of the point of observation -
the effects of all the other vortices in a random tangle
canceling each other out. 
Fig.~\ref{fig9} is therefore consistent with the random flow 
interpretation which results
from the analysis of the correlation functions.
It is also worth noticing that the $E(k) \sim k^{-3}$ scaling which 
appears in the
region $k_a<k<k_{\xi}$ is characteristic of the vortex core
\cite{KrstulovicBrachet2010,BradleyAnderson2012}.

\begin{figure}[t!]
	\centering{\includegraphics[scale = 0.45]{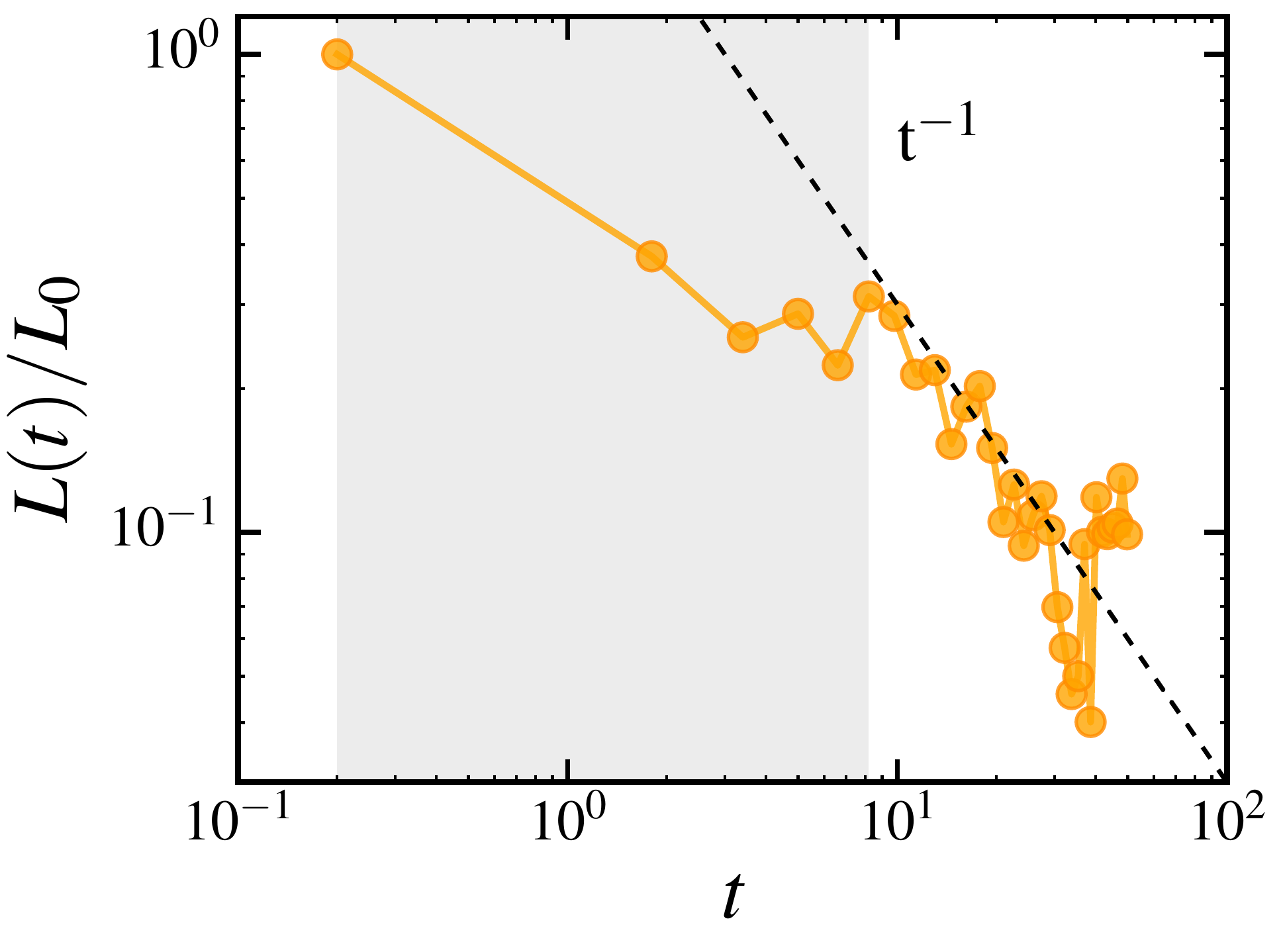}}\\
	\caption{
		(Color online). 
		Time evolution of vortex line density ($t$ in units of $\tau_{\ho}$). The shaded area corresponds to the 
		time elapsed for the decay of the two antiparallel multiply-charged
		vortices,
		$t\approx 8~\tau_{\ho}$, after which we only see singly quantized vortices in the system in the quasi-isotropic state.}
	\label{fig10}
\end{figure}


Besides the correlation function and the energy spectrum,
further insight into the nature of our turbulence
is acquired by measuring the temporal decay of the
vortex line density $L(t)$. This decay is caused by sound radiated away
by vortices as they accelerate about each other \cite{Leadbeater2003} or 
reconnect \cite{Leadbeater2001} with each other. 
Fig.~\ref{fig8} shows that, at large $t$, 
the decay is consistent with the form $L(t) \sim t^{-1}$,
as reported for a larger spherical condensate \cite{White2010};
the decaying turbulence is shown in the second movie in
the Supplementary Material \cite{SuppMovie}.
The $L(t) \sim t^{-1}$ decay is characteristic
of the `Vinen' turbulence regime (also called `ultra-quantum regime')
which was generated in superfluid helium by Walmsley \& Golov
\cite{Walmsley2008a} by short injections of vortex rings; 
long injections generated a `Kolmogorov' (or `quasi-classical')
regime which decayed as $L(t) \sim t^{-3/2}$. Numerical simulations 
\cite{Baggaley-ultra} of 
Walmsley \& Golov's experiment revealed that
(at large $k$ within the hydrodynamical range $k<k_{\ell}$)
`Vinen turbulence' (decaying as $L(t) \sim t^{-1}$) has
spectrum $E(k) \sim k^{-1}$ and `Kolmogorov turbulence' 
(decaying as $L(t) \sim t^{-3/2}$) has spectrum $E(k) \sim k^{-5/3}$.

The distinction between `Vinen turbulence' 
and `Kolmogorov turbulence', first
proposed by Volovik \cite{Volovik2003}, seems to be
characteristic of superfluids. It is now understood that Vinen's
pioneering experiments \cite{Vinen1957}
on counterflow heat transfer in superfluid
helium generated `Vinen turbulence': this was demonstrated by numerical
models of counterflow turbulence
\cite{Sherwin2012} which produced the expected $E(k) \sim k^{-1}$
energy spectrum and the $L(t) \sim t^{-1}$ vortex line density decay,
which is the large-$t$ decaying solution of the equation 
$dL/dt \sim -L^2$ proposed by Vinen on simple
physical arguments to model a random-like flow. 

A recent work \cite{Stagg2016} has examined the properties of
the turbulence following a thermal quench of a Bose gas. It found that
topological defects created by the Kibble-Zurek mechanism evolve 
into a turbulent vortex tangle \cite{BerloffSvistunov2002}
which eventually decays into a 
vortex free state. During the decay, which has the form $L(t) \sim t^{-1}$,
the energy spectrum is $E(k) \sim k^{-1}$ and the correlation functions
drop to few percent over the distance $\ell$. The thermal quench is therefore
another clear example of `Vinen turbulence'.

\section{Conclusion}

In this work we have explored the decay of initially imprinted
multicharged vortices as a method to generate turbulence in a trapped
Bose-Einstein condensate which is relatively free from large surface
oscillations and fragmentation.
We have examined the decay of two multicharged vortex systems,
in a typical cigar-shaped harmonically confined atomic Bose-Einstein condensate. 
The first (a quadruply charged vortex) led to the disordered, anisotropic vortex state. The second (two antiparallel doubly-charged
vortices) generated helical Kelvin waves on opposite oriented vortex lines 
which reconnected, creating a second disordered, quasi-isotropic vortex state. 
Looking for similarities with ordinary turbulence in its simplest
possible forms - in particular with the property of isotropy - we have 
concentrated the attention on the quasi-isotropic state, and 
carefully considered in which sense it is turbulent.

This question is subtle. 
In classical physics, 
turbulence implies a large range of length scales which are all excited
and interact nonlinearly. 
In ordinary fluids, the presence and the intensity of turbulence is 
inferred from the Reynolds number $\rm Re$, which must be sufficiently large
(typically few thousands, depending on the problem) for turbulence to exist.
But the Reynolds number has two definitions.  The first definition is
\begin{equation}
{\rm Re}=\frac{uD}{\nu},
\label{eq:Re1}
\end{equation}
\noindent
where $D$ is the system's large length scale, i.e. the system's size or 
the size of the
energy containing eddies, $u$ is the flow's
velocity at that large scale, and $\nu$ is the kinematic viscosity; 
this definition follows directly from the Navier-Stokes equation, and measures
the ratio of the magnitudes of inertial and viscous forces 
acting on a fluid parcel. The definition makes apparent why large
scale (e.g. geophysical) flows are always turbulent, and why microfluids
flows are not (indeed, with microfluids one has to rely on chaos, not 
turbulence, for achieving any desired mixing).
The second definition assumes Kolmogorov theory, and is
\begin{equation}
{\rm Re}=\left( \frac{D}{\eta} \right) ^{4/3}
\label{eq:Re2}
\end{equation}
\noindent
(where $\eta$ is the length scale of viscous dissipation). This definition
measures the degree of separation between the large length scale (at which
energy is typically injected) and the small length scale (at which
energy is dissipated).
In the context of condensates, the two definitions clash with each other:
The first definition implies that $\rm Re$ is infinite 
(because the viscosity is zero), and the second that $\rm Re$ is not
much larger than unity
(because the size of a typical condensate is larger, but not orders
of magnitude larger, than the healing length, which can be considered
the length scale at which acoustic dissipation of kinetic energy
occurs). Although interesting work is in progress to identify a 
definition of Reynolds number suitable for a superfluid system 
(for example exploring
dynamical similarities \cite{Reeves2015}), to answer the question which
we asked, at this stage, we have to leave the Reynolds number
and proceed in other ways.

We have therefore carefully examined the properties of the
disordered, quasi-isotropic state in terms of velocity statistics, energy spectrum,
correlation function and temporal decay, and compared them to the
properties of ordinary turbulence and of turbulent superfluid helium.
Clearly, the quasi-isotropic state does not compare well with the properties of ordinary turbulence. Despite the limited range of length scales available in a small BEC, we conclude that, in the decay of the two antiparallel doubly-charged vortices, the quasi-isotropy of our disordered state, the short correlation length, 
the $E(k) \sim k^{-1}$ scaling of the energy spectrum at the hydrodynamical
length scales,
and the $L(t) \sim t^{-1}$ temporal behavior of the decay,
identify our disordered, quasi-isotropic vortex state as an example
of the `Vinen turbulent regime' discovered in superfluid helium
at low temperatures.

The nature of the disorder, or turbulence,
in the anisotropic case generated by the decay of the single quadruply-charged 
vortex will be the subject of future investigations:
one should vary the amount of polarization and study
fluctuations of the velocity field over the mean rotating flow. 
There are no numerical studies yet of such turbulence in trapped Bose systems,
and (because of the role played by boundaries in the spin-down of 
viscous flows) no immediate classical analogies, so this case 
is less straightforward to analyze; 
spin-down dynamics experiments in superfluid helium
\cite{Walmsley2008b,Hosio2012a, Hosio2012b} should be the main 
reference systems.

Future work will also address the next
natural question: can the classical Kolmogorov
regime be achieved in much larger condensates under suitable forcing
at the largest length scale, as suggested by 
some numerical simulations \cite{Kobayashi2007},
thus identifying the cross over between Vinen and Kolmogorov turbulence?

\section*{Acknowledgements}

We acknowledge financial support from CAPES (PDSE Proc. No. BEX
9637/14-1), CNPq, FAPESP (program CEPID), and EPSRC grant EP/I019413/1. This research was developed making use of the computational resources (Euler cluster) from the Center for Mathematical Sciences Applied to Industry (Centro de Ci\^{e}ncias Matem\'{a}ticas Aplicadas \`{a} Ind\'{u}stria - CeMEAI) financed by FAPESP.

\bibliography{bib}

\begin{thebibliography}{72}
\bibitem [{gPD()}]{gPDF}%
  \BibitemOpen
  \href@noop {} {}\bibinfo {note} {The definition is
  $\rm{gPDF}(v_i)=(2\pi\sigma^2_i)^{-1/2}~ \exp{[-(v_i-\mu_i)^2/(2
  \sigma_i^2)]}$ where $\mu_i$ and $\sigma_i$ ($i=x,y,z$) are the mean values
  and the standard deviations respectively.}\BibitemShut {Stop}%
\end{thebibliography}
\end{document}